\documentstyle[12pt]{article}
\begin{document}
\begin{titlepage}

\centerline{\bf BFKL Pomeron at non-zero temperature} 

\centerline{\bf and integrability of the Reggeon dynamics in multi-colour QCD}

\date{\today}

\begin{center}
H. de Vega \\
Laboratoire de Physique Th\'eorique et Hautes Energies,\\
Universt\'e Pierre et Marie Curie, Paris VI et Universt\'e Denis
Diderot, Paris VII, BP 126,\\
4, place Jussieu, F-75252 Paris cedex 05, France.\\
Laboratoire Associ\'e au CNRS UMR 7589.\\
\end{center}
\begin{center}
L.N. Lipatov {$^{*}$\\
Theoretical Physics Department,\\
Petersburg Nuclear Physics Institute,\\
Gatchina, 188300, St. Petersburg, Russia}
\end{center}
\vskip 15.0pt

\centerline{\bf Abstract}

\noindent
We consider the QCD scattering amplitudes at high energies $\sqrt{s}$
and fixed momentum transfers $\sqrt{-t}$ in the leading logarithmic
approximation at a non-zero temperature $T$ in the $t$-channel.  It is
shown that the BFKL Hamiltonian has the property of holomorphic
separability. The Pomeron wave function for arbitrary $T$ is calculated
using an integral of motion. In multi-colour QCD, the
holomorphic Hamiltonian for $n$-reggeized gluons at temperature $T$ is
shown to coincide with the local Hamiltonian of an integrable
Heisenberg model and can be obtained from the $T=0$ Hamiltonian by
an unitary transformation.
We discuss the wave functions and the spectrum of intercepts for the
colourless reggeon states.  \vskip 3cm

\hrule

\vskip 3cm

\noindent

\noindent $^{*}$ {\it Supported by the grant
INTAS 2000-366} \vfill

\end{titlepage}

1. In QCD the scattering amplitudes $A(s,t)$ in Regge kinematics for high
energies $2E=\sqrt{s}$ and fixed momentum transfers $q=\sqrt{-t}$ are
obtained in the leading logarithmic approximation $\alpha _{s}\ln s\sim
1,\,\alpha _{s}=\frac{g^{2}}{4\pi }\rightarrow 0$ ($g$ is the QCD coupling
constant) by summing the largest contributions $\sim \left( \alpha _{s}\ln
s \right) ^{n}$ to all orders of perturbation theory within
the approach of Balitsky, Fadin, Kuraev and Lipatov (BFKL) \cite{BFKL}. The
BFKL Pomeron in the $t$-channel turns out to be a composite state of two
reggeized gluons (it is valid also in the next-to-leading approximation
\cite{NLO}). Its wave function $\Psi (\vec{\rho}_{1},\vec{\rho}_{2})$ satisfies
the Schr\"{o}dinger equation in the two dimensional impact-parameter space
$\vec{\rho}$
\begin{equation}
E\,\Psi (\vec{\rho}_{1},\vec{\rho}_{2})=H_{12}\,
\Psi (\vec{\rho}_{1},\vec{\rho}_{2})\;.
\end{equation}
The intercept $\Delta $ of the Pomeron, related to the high energy
asymptotics $\sigma _{t}\sim s^{\Delta }$ of the total cross-section, is
proportional to the ground state energy $E$
\[
\Delta =-\frac{\alpha _{s}N_{c}}{2\pi }\,E\;.
\]
The kinetic part $H_{kin}=\ln \,|p_{1}|^{2}+\ln \,|p_{2}|^{2}$ of $H_{12}$
is a sum of two gluon Regge trajectories and its potential part $H_{pot}$ is
related by a similarity transformation to the two-dimensional Green
function $\ln \,|\rho _{12}|^{2}$, where $\rho _{12}=\rho _{1}-\rho _{2}$.
[We introduced here the complex coordinates $\rho _{r}=x_{r}+iy_{r}$ and the
corresponding momenta $p_{r}=i\partial _{r}$].

The BFKL equation is used for the description of the deep-inelastic
lepton-hadron scattering together with the DGLAP equation \cite{DGLAP} (see
for example \cite{report}). It is invariant under the M\"{o}bius
transformations \cite{conf}
\[
\rho _{r}\rightarrow \frac{a\rho _{r}+b\rho _{r}}{c\rho _{r}+d\rho _{r}}
\]
with arbitrary complex parameters $a,b,c,d$ and $H_{12}$ has the property of
holomorphic separability (see \cite{report} \cite{holom})
\begin{equation}\label{HT0}
H_{12}=h_{12}+h_{12}^{\ast }\,,\,\,h_{12}=\sum_{r=1}^{2}\left[ \ln \,p_{r}+
\frac{1}{p_{r}}\,\ln \,(\rho _{12})\,\,p_{r}-\psi (1)\right] \;  ,
\label{hamil}
\end{equation}
where $\psi (z)=\Gamma ^{\prime }(z)/\Gamma (z)$.

The wave functions $\Psi $ belong to the principal series of unitary
representations of the M\"{o}bius group with conformal weights $
m=1/2+i\nu +n/2,\,\,\widetilde{m}=1/2+i\nu -n/2$ expressed in terms of the
anomalous dimension $\gamma =1+2i\nu $ and the integer conformal spin $n$
for the local gauge-invariant operators \cite{conf}. The conformal weights
are related to the eigenvalues $m(m-1)$, $\widetilde{m}(\widetilde{m}-1)$ of
the Casimir operators $M^{2}$ and $M^{2\ast }$, where
\[
M^{2}=\left( \sum_{r=1}^{2}M_{3}^{(r)}\right) ^{2}+\frac{1}{2}\left(
\sum_{r=1}^{2}M_{+}^{(r)}\,\sum_{s=1}^{2}M_{-}^{(s)}+
\sum_{r=1}^{2}M_{-}^{(r)}\,\sum_{s=1}^{2}M_{+}^{(s)}\right) =\rho
_{12}^{2}\; p_{1}\; p_{2}\;.
\]
Here $\vec{M}^{(r)}$ are the M\"{obius} group generators
\[
M_{3}^{(r)}=\rho _{r} \, \partial _{r}\,,\,\,\,M_{+}^{(r)}=\partial
_{r}\,,\,\,\,M_{-}^{(r)}=-\rho _{r}^{2} \; \partial _{r}
\]
and $\partial _{r}=\partial /\partial \rho _{r}$.

The eigenfunctions of $H_{12}$ can be considered as the three-point
functions of a two-dimensional conformal field theory and have the property of
holomorphic factorization \cite{conf},
\begin{equation}\label{fmm}
f_{m,\widetilde{m}}(\overrightarrow{\rho _{1}},\overrightarrow{\rho _{2}};
\overrightarrow{\rho _{0}})=\left\langle 0\right| \,\varphi
(\overrightarrow{
\rho _{1}})\,\varphi (\overrightarrow{\rho _{1}})\,O_{m\widetilde{,m}}(
\overrightarrow{\rho _{0}})\,\left| 0\right\rangle =\left( \frac{\rho
_{12}}{\rho _{10}\,\rho _{20}}\right) ^{m}\left( \frac{\rho _{12}^{\ast }}{\rho
_{10}^{\ast }\,\rho _{20}^{\ast }}\right) ^{\widetilde{m}}\,.
\end{equation}
One can calculate the energy putting this Ansatz in the BFKL
equation\cite{BFKL}
\begin{equation}
E_{m,\widetilde{m}}=\varepsilon _{m}+\varepsilon _{\widetilde{m}}
\quad , \quad \varepsilon _{m}=\psi (m)+\psi (1-m)-2\psi (1)\; .
\end{equation}
The minimum of $E_{m,\widetilde{m}}$ is obtained at $m=\widetilde{m}=1/2$
leading to a large intercept $\Delta =4 \; \frac{\alpha _{s}}{\pi
}N_{c} \, \ln 2$ of the BFKL Pomeron. In the next-to-leading
approximation the intercept is comparatively small ($\Delta \sim 0.2$
for the QCD case) \cite{photons}.

2. On the other hand, now a significant interest is devoted to the
quark-gluon plasma (QGP) generation in heavy nucleus
collisions (see for example \cite{QGP}). Current theoretical
understanding suggests that the QGP thermalizes via parton-parton
scattering. The QGP is understood to  cool down by hydrodynamic
expansion till the temperature reaches the hadronization scale $\sim 160$MeV.
One interesting phenomena is the suppression of the $\psi $-meson
production in the heavy nucleus collisions due to the
disappearance of the confining potential between $q$ and $\bar{q}$ at
high temperature \cite{QGP}. 
A similar effect should exist for glueballs
constructed from gluons. Because the Pomeron is considered as a composite
state of reggeized gluons, the influence of the temperature on its
properties is of great interest. In this paper we construct the BFKL
equation at temperature $T$ in the center-mass system of the $t$
-channel (where $\sqrt{t}=2\epsilon $) and investigate the integrability
properties of the BFKL dynamics in a thermostat for composite states of $n$
reggeized gluons in multi-colour QCD.

Let us consider the Regge kinematics in which the total particle
energy $\sqrt{s}$ is asymptotically large in comparison with the
temperature $T$. In this case one can neglect
the temperature effects in the propagators of the initial and intermediate
particles in the direct channels $s$ and $u$. But the momentum transfer
$|q|$ is considered to be of the order of $T$ (note, that $q_{\mu }$
is the vector orthogonal to the initial momenta $q_{\mu }\approx
q_{\mu }^{\perp }$ ). As it is well known \cite{T}, the particle wave
functions $\psi (x_{\mu })$ at temperature $T$ are periodic in the
euclidean time $\tau =i\,t$ with period $1/T$.

We introduce the temperature $T$ in the center of mass frame of the
$t$-channel. Thus, the euclidean energies of the intermediate
gluons in the $t$-channel become quantized as
$$
k_{4}^{(l)}=2\pi l\,T \; .
$$
In the $s$-channel the
invariant $t$ is negative and therefore the analytically continued 4-momenta
of the $t$-channel particles can be considered as euclidean vectors. It
means, that at temperature $T$, the wave functions
for virtual gluons are periodic functions of the 
holomorphic impact-parameter
$\rho =x+iy$ 
with imaginary period $\frac{i}{T}$. Also,
the canonically conjugated momenta $p$ have their imaginary part quantized,
\begin{equation}
\rho \rightarrow \rho +\frac{i}{T} \quad , \quad p=\mbox{Re}\, p+ 
\pi\, i \, l \, T \; .
\end{equation}
with integer $l$ (note that $p=(p_1+ip_2)/2 $).

It is convenient to rescale the transverse coordinates and corresponding
momenta as follows
\[
\rho \rightarrow \frac{1}{2\pi \,T} \; \rho \quad , \quad
p\rightarrow 2\pi \,T \, p \;.
\]
In these dimensionless variables one obtains
\[
0<\mbox{Im} \, \rho <2\pi \quad , \quad \mbox{Im} \, p= \frac{l}{2} \,.
\]
The calculation of the Regge trajectory $1+\omega (t)$ of the gluon 
at temperature $T$ in the t-channel,
in one-loop approximation reduces to the integration over the
real part $k_{1}$ of the transverse momentum $\vec{k}_{\perp}$
of the virtual gluon and to the summation over its imaginary part
$k_{2}=l$. In such a way we obtain
the following result for the trajectory having the separability
property [cf. \cite{holom}],
\[
\omega (-\vec{q}^{2})=-\frac{g^{2}}{8\pi ^{2}} \; N_{c} \; \Omega (-\vec{q}
^{2})\quad , \quad  \Omega (-\vec{q}^{2})=\Omega (q)+\Omega (q^{\ast }) \; .
\]
Here,
\[
\Omega (q)=\frac{\pi T}{\lambda }+\frac{1}{2}\,\left[ \psi (1+iq)+\psi
(1-iq)-2\psi (1)\right] \; ,
\]
where we regularized the infrared divergence for the zero mode $l=0$
introducing a mass $\lambda $ for the gluon (see \cite{BFKL}).

A similar  divergence appears in the Fourier transformation $G(
\overrightarrow{\rho }_{12})$ of the effective gluon propagator $(\vec{k}
_{\perp }^{2}+\lambda ^{2})^{-1}$ contained in the product of the effective
vertices $q_{1}k^{-1}q_{2}^{\ast }$ for the production of a gluon
with momentum $k_{\mu }$ (cf. \cite{report})
\[
G(\overrightarrow{\rho }_{12})=-\frac{\pi \,T}{\lambda }+\ln \left( 2\sinh
\frac{\rho _{12}}{2}\right) +\ln \left( 2\sinh \frac{\rho _{12}^{\ast }}{2}
\right) \; .
\]
Therefore, the divergence at $\lambda \rightarrow 0$ cancels in the
sum of kinetic and potential contributions to the BFKL equation and the
Hamiltonian $H_{12}$ for the Pomeron in a thermostat has the property of
holomorphic separability with the holomorphic Hamiltonian given below [cf.
eq.(\ref{hamil})]
\begin{equation}
h_{12}=\sum_{r=1}^{2}\left[ \frac12 \; \psi (1+ip_{r})+
\frac12 \; \psi(1-ip_{r}) +
\frac{1}{p_{r}}\,\ln \left( 2\sinh \frac{\rho_{12}}{2}\right) \,p_{r}-
\psi(1)\right] \; .  \label{hamilT}
\end{equation}

3. The Hamiltonian (\ref{hamilT}) is a periodic function of $\rho _{12}$.
Therefore its eigenfunctions are quasi-periodic functions of this
variable. The behaviour  of $h_{12}$ for small $\rho _{12}$
corresponds to
the low temperature regime. Hence, the eigenfunctions of $h_{12}$
behave for $  \rho_{12} \to 0 $ as the holomorphic part of the
zero-temperature wave functions eq.(\ref{fmm}),
\begin{equation}\label{ro0}
\Psi_m (\rho _{12}) \buildrel{ \rho_{12} \to 0}\over= \rho_{12}^{m}
\end{equation}
Notice that $ \Psi_{1-m} (\rho _{12})$ is an eigenfunction too.

We find the small-$T$ expansion of $h_{12}$ near its singularities $\rho
_{12}=2\pi il$. For example, for small $\rho _{12}$ and large
$p_{1},\,p_{2}$ we have
\[
h_{12} = h_{12}^{0}+\sum_{k=1}^{\infty} \frac{B_{2k}}{2k} \sum_{r=1,2}\left[
\frac{(-1)^{k+1}}{p_{r}^{2k}}+\frac{1}{p_{r}}\,
\frac{\rho_{12}^{2k}}{(2k)!} \,\,p_{r}\right] \; ,
\]
where $h_{12}^{0}$ is the holomorphic BFKL Hamiltonian at a zero
temperature [given by eq.(\ref{HT0})] and the $B_{2k}$ are the
Bernoulli numbers. This
representation for $h_{12}$ permits us to find the small-$T$ expansion of its
eigenfunctions. For example, at a vanishing momentum transfer $Q=p_{1}+p_{2}$
we obtain for the eigenfunction $\Psi_m (\rho _{12})$,
\[
\Psi_m (\rho _{12})=\rho_{12}^{m}\left[
  1-\frac{1}{24}\frac{m(m-1)}{2m+1} \;
  \rho_{12}^{2}+\frac{1}{5760}\frac{m(m-1)(5m^{2}+7m+6)}{(2m+1)(2m+3)}
  \; \rho_{12}^{4}+ {\cal O}(\rho_{12}^6) \right] .
\]
These eigenfunctions are parametrized by the conformal weight $m$.

On the other hand, from the above expansion it is possible to verify, that
the holomorphic Hamiltonian has the non-trivial integral of motion:
\begin{equation}\label{A}
A=4\sinh ^{2}\frac{\rho _{12}}{2}\,\,p_{1}\,p_{2}\quad , \quad
[A,h_{12}]=0\,.
\end{equation}
Therefore, instead of solving the Schr\"{o}dinger equation we can
search for the  eigenfunctions of the operator $A$. For non-zero $Q$
one can write the holomorphic wave function as a product of a plane wave
depending on $R=(\rho _1+\rho_2)/2$ times a
solution of the following equation for the relative motion of two gluons
\[
\left[ \frac{Q^{2}}{4}+\frac{\partial ^{2}}{\partial \rho^2}\right]
\; \Psi (\rho ,Q)=\frac{m(m-1)}{4 \sinh ^{2}\frac{\rho}{2}}\; \;\Psi (\rho ,Q)
\quad , \quad
\rho =\rho_{12}\,,\,\,t=-4\left| Q\right| ^{2} \; .
\]
The two independent solutions of the above differential equation can be
expressed in terms of hypergeometric functions
\begin{equation}
\Psi_{1}^{(m)}(\rho ,\,Q)=e^{\frac{i}{2} Q \, \rho} \; \left(e^{\rho}
- 1 \right)^m \; F\left(iQ+m,m;2m;1 - e^{\rho}\right)
\quad , \quad \Psi_{2}^{(m)}(\rho ,\,Q) \equiv \Psi _{1}^{(1-m)}(\rho
,\,Q) \; .
\end{equation}
For $\rho \rightarrow 0$ eq.(\ref{ro0}) holds and the singularities of
$\Psi^{(r)}(\rho ,\,Q)$ at $1 - e^{\rho}=1$ and $ 1 - e^{\rho}=\infty
$ correspond to the points $\rho =-\infty $ and $\rho =\infty $, respectively.

The analytic continuation of $\Psi ^{(r)}$ along the imaginary axes from $
\rho =0$ to $\rho =2\pi i$ is equivalent to the continuation of these
eigenfunctions in a circle passed in a clock-wise direction around the
singularity at $\rho =-\infty $. The monodromy matrix expressing the
analytically continued  solutions in terms of the initial ones can be
easily calculated.

The Pomeron wave functions can be written as a
bilinear combination of  holomorphic and anti-holomorphic eigenfunctions $\Psi
^{(r)}(\rho ,\,Q)$ and $\Psi^{(r)}(\rho^{\ast } ,\,Q^{\ast })$. The
property of single-valuedness in the cylinder topology
corresponding to the periodicity on the boundaries of the strip
$0<\mbox{Im} \,  \rho_{12}<2\pi $ is easily imposed to such Pomeron
wave function using  the monodromy matrix for $\Psi ^{(r)}(\rho
,\,Q)$. The resulting wave function can be written as,
\begin{equation}\label{hip}
\Psi ^{(m,\widetilde{m})}(\vec{\rho},\,\vec{Q})=\chi _{1}^{(m)}(\rho
,\,Q)\; \chi _{1}^{(\widetilde{m})}
(\rho^{\ast },\,Q^{\ast }) - (-1)^N
\; \chi _{2}^{(m)}(\rho ,\,Q)\;\chi _{2}^{(\widetilde{m})}(\rho ^{\ast
},\,Q^{\ast })\,,
\end{equation}
where,
$$
\chi _{1}^{(m)}(\rho ,\,Q) = 2^{1-2\,m} \;
\frac{\Gamma\left(m+iQ\right)}{\Gamma\left(m+\frac12\right)} \;
\Psi_{1}^{(m)}(\rho ,\,Q) \; , \; \chi _{2}^{(m)}(\rho ,\,Q) =
\chi_{1}^{(1-m)}(\rho ,\,Q) 
$$
and $ N = 2 \, \mbox{Im} \, Q $ is an integer.

4. The Pomeron wave function can be constructed directly in coordinate
space. For this purpose we use the conformal transformation
\begin{equation}\label{exp}
\rho _{r}=\ln \,\rho _{r}^{\prime }
\end{equation}
and the integral of motion eq.(\ref{A}) becomes
\[
A=-(\rho _{12}^{\prime })^{2}\frac{\partial }{\partial \rho
_{1}^{\prime }}\, \frac{\partial }{\partial \rho _{2}^{\prime }}\,.
\]
Thus, $A$ coincides in the variables $\rho _{r}^{\prime }$
with the Casimir operator of the conformal group whose eigenfunctions are
well known (see \cite{conf}). Thus, the Pomeron wave function at non-zero
temperature having the property of single-valuedness and periodicity
takes the form
\begin{equation}
\Psi ^{(m,\widetilde{m})}(\vec{\rho _{1}},\,\vec{\rho _{2}},\,\vec{\rho
_{0}})=\left( \frac{\sinh \frac{\rho _{12}}{2}}{2\sinh \frac{\rho _{10}}{2}
\,\sinh \frac{\rho _{20}}{2}}\right) ^{m}\,\left( \frac{\sinh \frac{\rho
_{12}^{\ast }}{2}}{2\sinh \frac{\rho _{10}^{\ast }}{2}\,\sinh \frac{\rho
_{20}^{\ast }}{2}}\right) ^{\widetilde{m}}\,.
\end{equation}
The orthogonality and completeness relations for these functions can be
easily obtained from the analogous results for $T=0$ (see \cite{conf})
using the above conformal transformation. These wave functions are 
proportional to the Fourier transformation
of the wave functions $\Psi ^{m,\widetilde{m}}(\vec{\rho},\vec{Q})$.

Moreover, the pair BFKL Hamiltonian $h_{12}$ can be expressed in terms of
the BFKL Hamiltonian at zero temperature in the new variables
\begin{equation}
h_{12}=\ln (p_{1}^{\prime }\,p_{2}^{\prime })+\frac{1}{p_{1}^{\prime }}
\,\log (\rho _{12}^{\prime })\,\,p_{1}^{\prime }+\frac{1}{p_{2}^{\prime }}
\,\log (\rho _{12}^{\prime })\,\,p_{2}^{\prime }-2\psi (1)\,,
\end{equation}
where $p_{r}^{\prime }=i\frac{\partial }{\partial \rho _{r}^{\prime
}}$. In the course of the derivation the 
following operator identity (see \cite{report})
\[
\frac{1}{2}\left[ \psi \left(1+z\frac{\partial }{\partial z}\right)+\psi
  \left(-z\frac{ \partial }{\partial z}\right)\right] =\ln \,z+\ln
  \,\frac{\partial }{\partial z}
\]
was used to transform the kinetic part as well as  properties of the
$\psi $-function.

In summary, the exponential mapping eq.(\ref{exp}) which in dimensional
variables takes the form
$$
\rho' = \frac{1}{2\pi \, T} \; e^{2\pi \, T \; \rho} \; ,
$$
maps the reggeon dynamics from zero temperature to temperature
$T$. This mapping explicitly exhibits a periodicity 
$\rho \rightarrow \rho +\frac{i}{T}$ for a thermal state. 
It must be noticed that such class
of mappings are known to describe thermal situations for
quantum fields in accelerated frames and in black hole backgrounds\cite{bh}.

\bigskip

5. As it is well known \cite{BKP}, the BFKL equation at $T=0$ can be
generalized to composite states of $n$ reggeized gluons. In
the multi-colour limit $N_{c}\rightarrow \infty $ the BKP equations are
significantly simplified thanks to their conformal invariance
\cite{conf},  holomorphic separability \cite{holom} and integrals of
motion \cite{intodd}. The generating function for the holomorphic
integrals of motion
coincides with the transfer matrix for an integrable lattice spin model
\cite{brown} \cite{integr}. The transfer matrix is the trace of the
monodromy matrix
\[
t(u)=L_{1}(u)\,L_{2}(u)...L_{n}(u)\,,
\]
satisfying the Yang-Baxter equations \cite{integr}. The integrability of the
$n$-reggeon dynamics in multi-colour QCD is valid also at non-zero
temperature $T$, where, according to the above arguments we should
take the $L$-operator in the form
\[
L_{k}=\left(
\begin{array}{cc}
u+p_{k} & e^{-\rho _{k}}\,p_{k} \\
-e^{\rho _{k}}\,p_{k} & u-p_{k}
\end{array}
\right) \,.
\]
In particular, the holomorphic Hamiltonian is the local Hamiltonian of the
integrable Heisenberg model with the spins being unitarily transformed
generators of the M\"{o}bius group (cf. \cite{Heis} \cite{FadKor})
\[
M_{k}=\partial _{k} \quad ,  \quad M_{+}=e^{-\rho _{k}}\,\partial
_{k} \quad ,  \quad M_{-}=-e^{\rho _{k}}\,\partial _{k}\; .
\]
Because the Hamiltonian at  non-zero temperature can be obtained by an
unitary transformation from the zero temperature Hamiltonian, the
spectrum of the intercepts
for multi-gluon states is the same as for zero temperature \cite{JW}-
\cite{dkm} and the wave functions of the composite states can be calculated
by the substitution $\rho _{k}\rightarrow e^{\rho _{k}}$. 

Furthermore, the non-linear Balitsky-Kovchegov equation \cite{BalKov}
can be generalized to the case of non-zero temperature as follows,
\begin{equation}
\frac{\partial N_{\vec{\rho}_{1},\vec{\rho}_{2}}}{\partial
  Y}=\bar{\alpha}
_{s}\int \frac{d^{2}\rho _{0}}{2\pi }\,\frac{\left| \sinh \frac{\rho
_{12}}{2}\right| ^{2}}{4\left| \sinh \frac{\rho _{10}}{2}\right|
  ^{2}\left| \sinh 
\frac{\rho _{20}}{2}\right| ^{2}}\,\left( N_{\vec{\rho}_{1},\vec{\rho}
_{0}}+N_{\vec{\rho}_{2},\vec{\rho}_{0}}-N_{\vec{\rho}_{1},\vec{\rho}_{2}}-N_
{\vec{\rho}_{1},\vec{\rho}_{0}}\,N_{\vec{\rho}_{2},\vec{\rho}_{0}}\right) \,,
\end{equation}
where $N_{\vec{\rho}_{1},\vec{\rho}_{2}}$ is the amplitude of finding \ a
dipole with the impact parameters $\vec{\rho}_{1}$ and $\vec{\rho}_{2}$ in a
hadron and the integration over $\rho _{0}$ is performed over the strip $0<
 \mbox{Im}\,\rho _{0}<2\pi $. Note, however, that in this equation  one
takes into account only
fan diagrams for the Pomeron interactions  
among all possible diagrams for reggeized gluons 
appearing in  the high energy effective action \cite{eff}.

We thank G.\thinspace S. Danilov and V. N. Velizhanin for helpful
discussions.

\end{document}